\documentclass[secnumarabic,amssymb, nobibnotes, aps, prd]{revtex4-2}

\usepackage[colorlinks=true,allcolors=blue]{hyperref}       
\usepackage{amsmath}
\usepackage{natbib}
\usepackage{adjustbox}
\usepackage{subfigure}
\usepackage{graphicx}
\usepackage{booktabs}       
\usepackage{amsfonts}       
\usepackage{nicefrac}       
\usepackage{microtype}
\usepackage{color}
\usepackage{textcomp}
\usepackage{gensymb}
\usepackage{cleveref}
\usepackage{multirow}
\usepackage{makecell}
\usepackage{array}
\newcolumntype{C}[1]{>{\centering\arraybackslash}m{#1}}

\begin{document}

\title[]{Rotating Kiselev Black Holes in $f(R,T)$ Gravity}
\author{Sushant~G.~Ghosh$^{a,\;b}$} \email{sghosh2@jmi.ac.in, sgghosh@gmail.com}
\author{Shafqat Ul Islam$^{a,\;b}$} \email{shafphy@gmail.com}
\author{Sunil D. Maharaj$^{b}$}\email{maharaj@ukzn.ac.za}
\affiliation{$^{a}$ Centre for Theoretical Physics, 
	Jamia Millia Islamia, New Delhi 110025, India}
\affiliation{$^{b}$ Astrophysics and Cosmology Research Unit, 
	School of Mathematics, Statistics and Computer Science, 
	University of KwaZulu-Natal, Private Bag 54001, Durban 4000, South Africa}
\date{\today}

\begin{abstract}
Exact solutions describing rotating black holes can provide significant opportunities for testing modified theories of gravity, which are motivated by the challenges posed by dark energy and dark matter. Starting with a spherical Kiselev black hole as a seed metric, we construct rotating Kiselev black holes within the $f(R,T)$ gravity framework using the revised Newman-Janis algorithm — the $f(R,T)$ gravity-motivated rotating Kiselev  black holes (FRKBH), which encompasses, as exceptional cases, Kerr ($K=0$) and Kerr-Newman ($K=Q^2$) black holes. These solutions give rise to distinct classes of black holes surrounded by fluids while considering specific values of the equation-of-state parameter, $w$, for viable choices for the $f(R,T)$ function.  From the parameter space or domain of existence of black holes defined by $a$ and $\gamma$ for FKRBH, we discover that when $a_1<a<a_2$, there is a critical value $\gamma=\gamma_E$ which corresponds to extreme value black holes portrayed by degenerate horizons. When $a<a_1$ ($a>a_2$), we encounter two distinct critical values $\gamma=\gamma_{E1}, \; \gamma_{E2}$ with $\gamma_{E1}>\gamma_{E2}$ (or $\gamma=\gamma_{E3},\; \gamma_{E4}$ with $\gamma_{E3}>\gamma_{E4}$. We delve into the horizon and global structure of FKRBH spacetimes and examine their dependence on parameters $w$ and $\gamma$. This exploration is motivated by the remarkable effects of $f(R,T)$ gravity, which gives rise to diverse and intricate spacetime structures within the domain where black holes exist.
\end{abstract}
\maketitle

\section{Introduction}\label{Sec-1}
The  $f(R)$ gravity theory undoubtedly explains the existence of a late-time cosmic acceleration of the Universe \cite{Carroll:2003wy}. Despite $f(R)$ gravity's potential advantages,  this theory can not explain various observational tests conducted in the solar system, such as the motion of planets \cite{Erickcek:2006vf,Capozziello:2007ms}. It also faces challenges in explaining the cosmic microwave background (CMB) tests, the effects of strong gravitational lensing, and phenomena on galactic scales \citep{Dolgov:2003px,Olmo:2005hc,Yang:2008wu,Dossett:2014oia,Campigotto:2016cmt,Xu:2017qxf}. Further, $f(R)$ gravity struggles to account for stable stellar configurations \citep{Briscese:2006xu,Kobayashi:2008tq,Babichev:2009fi}. These shortcomings have prompted scientists to further develop and generalize the theory by exploring the coupling between scalar curvature and matter \citep{Nojiri:2004bi,Allemandi:2005qs,Bertolami:2007gv}. By incorporating this coupling, researchers hope to overcome the limitations and improve the overall consistency of the theory. Harko et al. \citep{Harko:2011kv} have put forward a generalized form of $f(R)$ gravity theory, namely $f(R,T)$ gravity.  The presence and distribution of matter affect the curvature of spacetime, and conversely, the curvature of spacetime influences the motion of matter. This coupling leads to exciting phenomena and potentially explains specific astrophysical observations. The $f(R,T)$ gravity is an intriguing approach that extends the Einstein field equations by introducing additional terms involving the curvature scalar $R$ and the trace of the energy-momentum tensor $T$. The theory received significant attention in recent years because of its potential to address various cosmological phenomena and serves as a modified theory of gravity that can account for dark matter and dark energy (see, e.g., \citep{SupernovaSearchTeam:1998fmf,SupernovaCosmologyProject:1997zqe,SupernovaCosmologyProject:1998vns,DeFelice:2010aj,Nojiri:2006ri,Nojiri:2010wj}). These articles provide an extensive overview of $f(R,T)$ gravity and discuss its theoretical foundations, mathematical formalism, and cosmological implications. They cover various topics, including models, observational constraints, and the impact of $f(R,T)$ gravity on the early and late universe. This gravitational theory's dependence on $T$ can be attributed to quantum effects, imperfect fluids, extra fluids, or an effective cosmological constant. The application of $f(R, T)$ gravity has produced exciting results in various regimes, as evidenced in references \citep{Das:2016mxq,Alvarenga:2013syu,Alves:2016iks,Yousaf:2016lls,Mubasher:2012mb}.

For high densities and pressures, the effects of modifying gravity through $f(R,T)$ theory are expected to become more pronounced. Therefore, it is natural to investigate the impact of these modifications on compact objects or black holes. In theories where the Lagrangian density depends on $T$, it is expected that there will be differences in the solutions compared to general relativity, particularly when considering non-zero energy momentum tensors. This implies additional effects arising from the coupling between matter and geometry. An intriguing system to explore effects is fluids surrounding spherical matter sources. Motivated by this,  Santos {\it et al.} \cite{Santos:2023fgd}
discussed spherical black hole solutions within the framework of $f(R,T)$ gravity, with the black hole being surrounded by the quintessence fluid discussed by Kiselev \cite{Kiselev:2002dx}. They investigated specific physical scenarios corresponding to solutions obtained by carefully selecting appropriate values for the fluid equation of state parameters. They solely focussed on viable choices for the $f(R,T)$ function to ensure that they could connect the obtained results to an extension of the Kiselev solution \cite{Kiselev:2002dx}.   However, testing astrophysical observations poses a challenge for spherical black hole models since black hole spin, i.e., rotating black holes commonly found in nature, significantly influences astrophysical processes \cite{Reynolds:2020jwt}. Furthermore, the lack of rotating black hole models in the modified gravity, including $f(R,T)$ gravity, hampers the ability to test the theory through observations \cite{Brahma:2020eos,Islam:2022wck,Afrin:2022ztr,Kumar:2020hgm,Kumar:2020owy}. The Kerr metric \cite{Kerr:1963ud}, which represents rotating black hole solutions in general relativity resulting from the collapse of massive stars, holds great significance.
Remarkably, the revised Newman-Janis method has been successful in generating rotating metrics from non-rotating seed metrics in other modified gravity models \cite{Brahma:2020eos, Liu:2020ola, Chen:2022nix, Ghosh:2014pba, Ghosh:2015ovj, Kumar:2020hgm, Kumar:2020owy, Kumar:2021cyl, Islam:2021dyk}. This success has motivated us to pursue a rotating or axisymmetric extension of the spherical metric obtained in \cite{Santos:2023fgd} or to find a Kerr-like metric, referred to as an $f(R,T)$ motivated rotating Kiselev black hole (FRKBH) metric, which can be tested using astrophysical observations. We employ the revised Newman-Janis algorithm starting from a spherical black hole in \cite{Santos:2023fgd} as a seed metric to construct a rotating spacetime. We examine the black hole's various properties, including its horizon's structure, and create Penrose and embedding diagrams for further analysis.
To methodically analyze the FRKBH, we use specific values for the fluid parameter $w$ corresponding to black holes surrounded by different fields. These fields include dust ($w=0$), radiation ($w=1/3$), quintessence ($w=-2/3$), and phantom ($w=-4/3$) \cite{Kiselev:2002dx,Ghosh:2015ovj}.

The paper is organized as follows: In Section \ref{sec2}, we review spherical symmetric black holes in $f(R,T)$ gravity. Then, in Section \ref{sec3}, we construct the rotating counterpart of the spherical seed metric (\ref{NR}), known as the FRKBH metric. Additionally, within the same section, we discuss the general features of the FRKBH metric, including horizon structures and energy conditions. Furthermore, we analyze the phase space of the FRKBH spacetimes, highlighting their unique properties in Section \ref{sec4}. In Section \ref{sec5}, we utilize the spacetime isometries to determine the conserved mass and angular momentum of the FRKBH spacetime. Finally, in Section \ref{sec6}, we summarise our main findings.

\section{Black Hole Solution}\label{sec2}
The $f(R,T)$ modified theory of gravity is considered to be  a generalisation of general relativity \cite{Harko:2011kv,Alves:2016iks,Alvarenga:2013syu,Yousaf:2016lls,Pinto:2022tlu,Myrzakulov:2012qp,Barrientos:2018cnx}.  The Einstein-Hilbert action in the context of $f(R,T)$ gravity takes the form 
\begin{equation}
S=\frac{1}{16\pi}\int f(R,T)\sqrt{-g}d^4x + \int L_{m}\sqrt{-g}d^4x,
    \label{e1}
\end{equation}
where the function $f(R,T)$ is an arbitary function of  the Ricci scalar $R$ and the trace $T$ of the energy-momentum tensor of matter $T_{\mu\nu}$. $L_m$ in Eq.~(\ref{e1}) denotes the matter Lagrangian density, which is linked to a specific energy-momentum tensor. By varying the action with respect to metric tensor $g_{\mu\nu}$ results \cite{Harko:2011kv}
\begin{align}\label{e2}
\delta S = & \frac{1}{16\pi}\int\left[f_{R}(R,T)R_{\mu\nu}\delta g^{\mu\nu} + f_{R}(R,T)g_{\mu\nu} \Box \delta  g^{\mu\nu}  \right.   
\nonumber\\
& -f_{R}(R,T)\nabla_{\mu}\nabla_{\nu}\delta g^{\mu\nu} + 
 f_{T}(R,T)\frac{\delta (g^{\eta\xi}T_{\eta\xi})}{\delta g^{\mu\nu}}\delta g^{\mu\nu}
 \nonumber\\
 & \left. -\frac{1}{2}g_{\mu\nu}f(R,T)\delta g^{\mu\nu}+\frac{16\pi}{\sqrt{-g}}\frac{\delta (\sqrt{-g}L_{m})}{\delta g^{\mu\nu}}\right]\sqrt{-g}d^4x, 
\end{align}
where $f_R(R,T)=\partial f(R,T)/\partial R$ and $f_T(R,T)=\partial f(R,T)/\partial T$. The variation of $T$ with respect to the metrics tensor yields 
\begin{equation} \label{e3}
\frac{\delta (g^{\eta\xi}T_{\eta\xi})}{\delta g^{\mu\nu}}=T_{\mu\nu} + \Theta_{\mu\nu}.   
\end{equation}
where
\begin{equation} \label{e4}
\Theta_{\mu\nu} \equiv g^{\eta\xi}\frac{\delta T_{\eta\xi}}{\delta g^{\mu\nu}}=-2T_{\mu\nu}+g_{\mu\nu}L_{m}-2g^{\eta\xi}\frac{\partial^2L_{m}}{\partial g^{\mu\nu}g^{\eta\xi}}.
\end{equation}
After integrating the second and third term in Eq.~(\ref{e2}), we obtain the field equations of the $f(R,T)$ gravity as
\begin{align}\label{e5}
 f_{R}&(R,T)R_{\mu\nu}-\frac{g_{\mu\nu}}{2}f(R,T)+
(g_{\mu\nu}\Box - \nabla_{\mu}\nabla_{\nu})f_{R}(R,T) \nonumber\\
&= 8\pi T_{\mu\nu} - f_{T}(R,T)T_{\mu\nu}- f_{T}(R,T)\Theta_{\mu\nu}\, .
\end{align}
When $f(R,T) \equiv f(R)$, Eq. (\ref{e5}) reduces to the field equations in the context of $f(R)$ gravity \cite{Ghosh:2015ovj}. The novel feature that $f(R,T)$ gravity introduces is the possibility of arbitrary coupling between matter and geometry. This paper considers a special case of $f(R,T)$ gravity such that the  $f(R,T)$ function is given by 
\begin{equation}
    f(R,T)=R + 2f(T),
    \label{e6}
\end{equation}
where $f(T)$ is an arbitrary function of the trace of the energy momentum tensor. On using the Eq.~(\ref{e6}) in Eq. (\ref{e5}), the field equations simplifies to
\begin{align}
 R_{\mu\nu}-\frac{g_{\mu\nu}}{2}R= & 8\pi T_{\mu\nu}- 2f'(T)T_{\mu\nu}
 \nonumber\\
 &- 2f_{T}'(T)\Theta_{\mu\nu} + f(T)g_{\mu\nu},
    \label{e7}
\end{align}
where $f'(T)=df(T)/dT$. 

We solve Eq.~(\ref{e7}) for the quintessence field introduced by Kiselev \cite{Kiselev:2002dx}, which is characterized by the equation of state $p=\omega \rho$. with $-1/3<\omega<-1$. One possible candidate to explain dark energy is quintessence field. The $T_{\mu\nu}$ for the quintessence matter reads \cite{Kiselev:2002dx}
\begin{align}
    T^{t}_{\:\:\:t}= T^{r}_{\:\:\:r}&=\rho(r), \label{e8}\\
   T^{\theta}_{\:\:\:\theta}= T^{\phi}_{\:\:\:\phi}&=-\frac{1}{2}\rho (3\omega+1),\label{e9}
\end{align}
and $\omega$ is the parameter of the equation of state. Kiselev black holes  \cite{Kiselev:2002dx} have the components of  energy-momentum tensor effectively connected to an anisotropic fluid  represented by 
\begin{equation}
    T^{\mu}_{\:\:\:\nu} = \mbox{diag}(\rho,-p_r,-p_t,-p_t),
    \label{e10}
\end{equation}
where $p_r=-\rho$ and $p_t=\frac{1}{2}\rho (3w+1)$, which can be extracted from the general form of the anisotropic fluid  \cite{Mota:2019opp}:
\begin{equation}
    T_{\mu\nu}=-p_{t}g_{\mu\nu}+(p_{t}+\rho)U_{\mu}U_{\nu}+(p_{r}-p_{t})N_{\mu}N_{\nu}
    \label{anytensor},
\end{equation}

\begin{figure*}
	\begin{centering}
		\begin{tabular}{p{9cm} p{9cm}}
		    \includegraphics[scale=0.65]{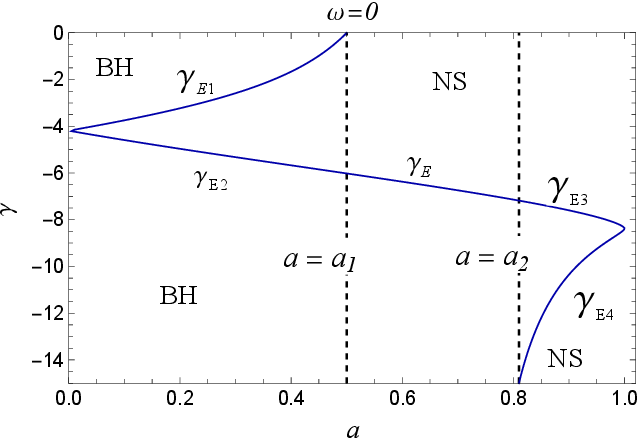}&
			\includegraphics[scale=0.6]{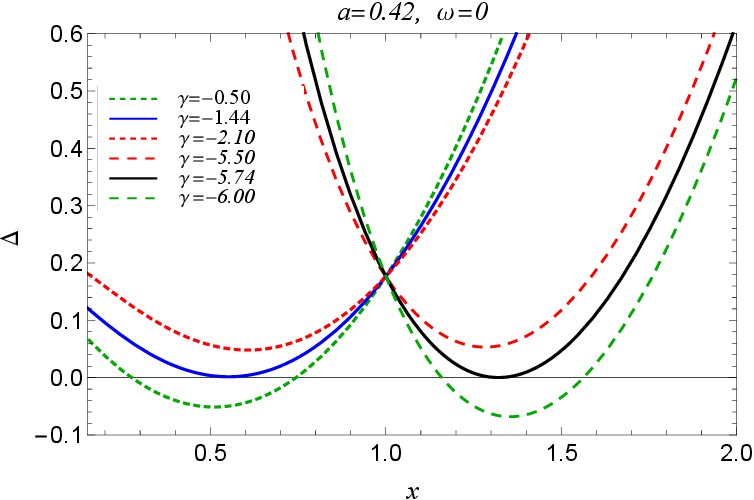}\\
			\includegraphics[scale=0.6]{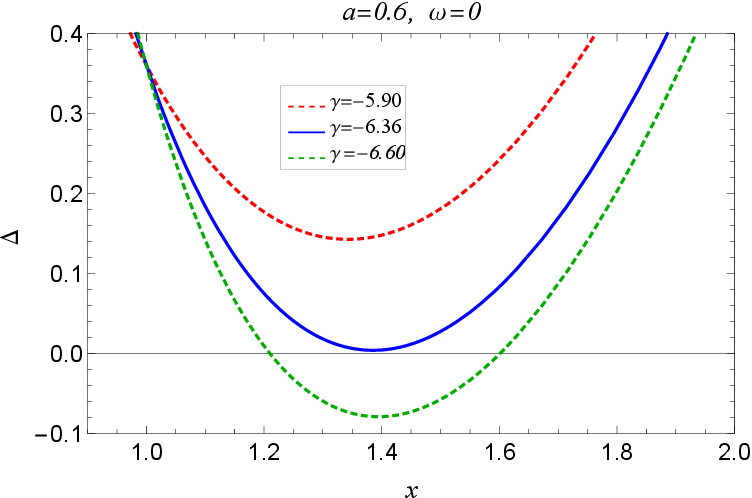}&
			\includegraphics[scale=0.6]{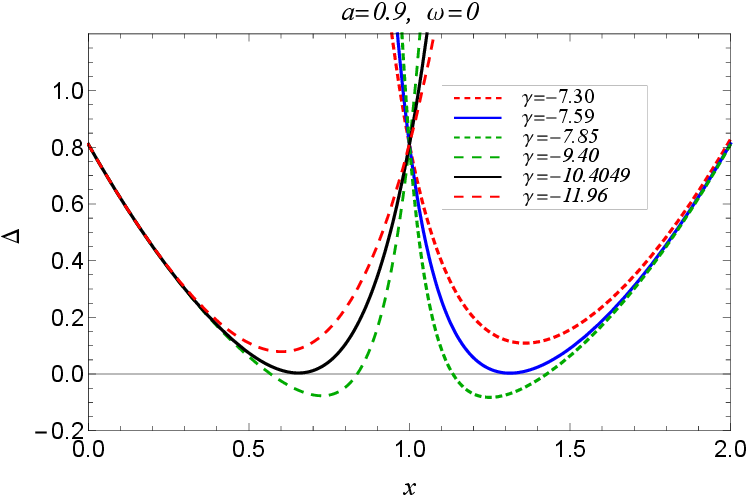}
		\end{tabular}
	\end{centering}
	\caption{The parameter space of $a$ and $\gamma$ for FKRBH. When $a_1<a<a_2$, an extreme value $\gamma=\gamma_E$ corresponding extremal value black holes exists. When $a<a_1$ ($a>a_2$), we have two extremal $\gamma=\gamma_{E1}$ and $\gamma=\gamma_{E2}$ with $\gamma_{E1}>\gamma_{E2}$ (or $\gamma=\gamma_{E3}$ and $\gamma=\gamma_{E4}$ with $\gamma_{E3}>\gamma_{E4}$). The behaviour of horizons at $\omega=0$ with varying black hole parameter $\gamma$ at (i) $a=0.42$ (top right), (ii) $a=0.60$ (bottom left), and (iii) $a=0.90$ (bottom right).}\label{plot1}	
\end{figure*}   

\begin{figure*}
	\begin{centering}
		\begin{tabular}{p{9cm} p{9cm}}
		    \includegraphics[scale=0.65]{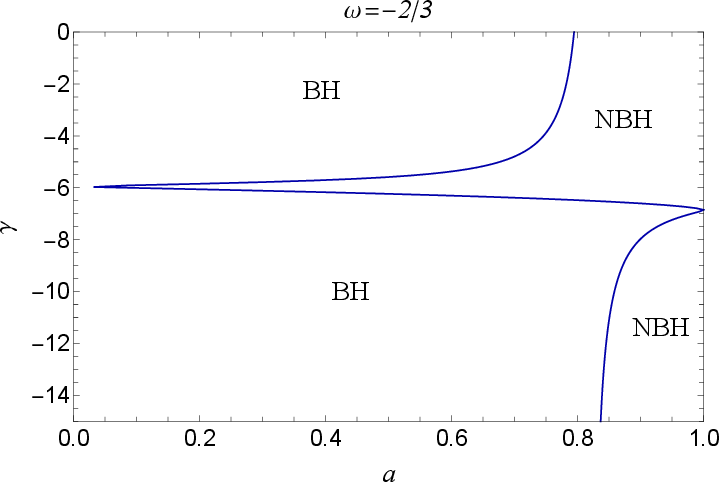}&
			\includegraphics[scale=0.6]{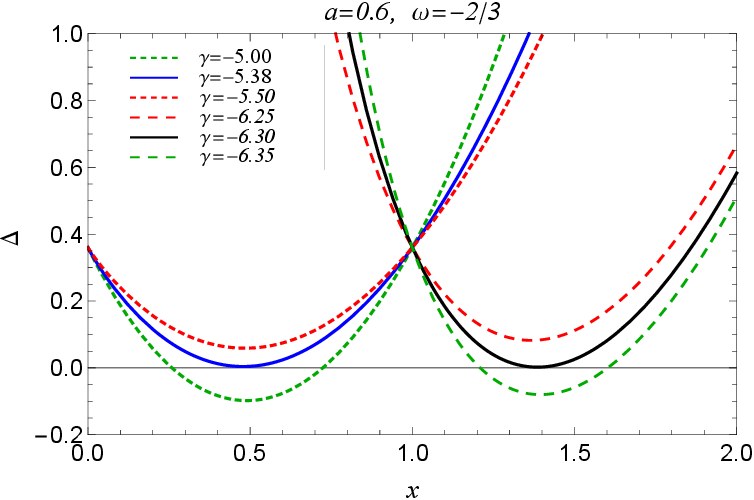}\\
			\includegraphics[scale=0.6]{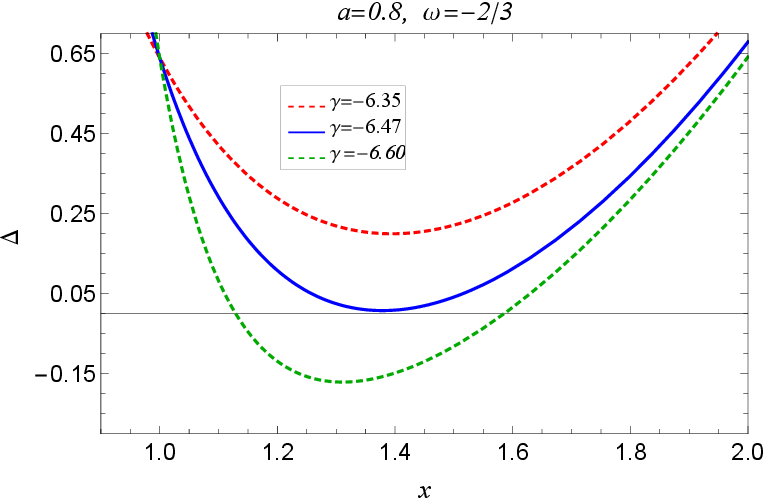}&
			\includegraphics[scale=0.6]{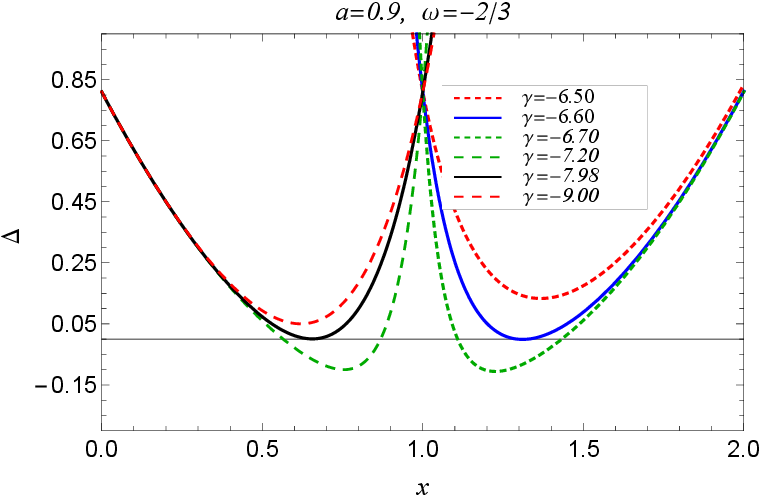}
		\end{tabular}
	\end{centering}
	\caption{The parameter space of $a$ and $\gamma$ at $\omega=-2/3$ (top right). The behaviour of horizons at $\omega=-2/3$ with varying black hole parameter $\gamma$ at (i) $a=0.60$ (top right) (ii) $a=0.80$ (bottom left) and (iii) $a=0.90$ (bottom right). The solid blue and black lines correspond to extreme values of parameters.}\label{plot2}		
\end{figure*} 

\begin{figure*}
	\begin{centering}
		\begin{tabular}{p{9cm} p{9cm}}
		    \includegraphics[scale=0.65]{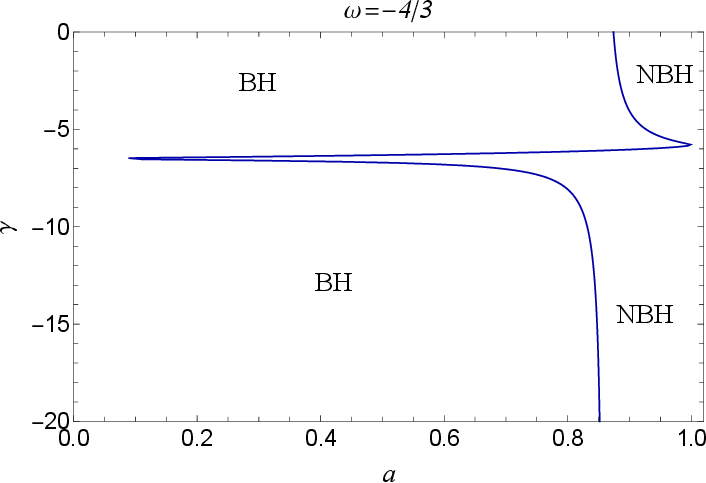}&
			\includegraphics[scale=0.6]{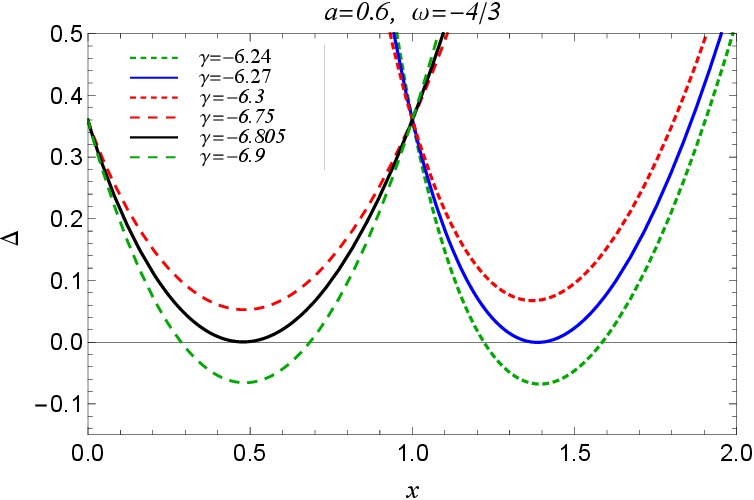}\\
			\includegraphics[scale=0.6]{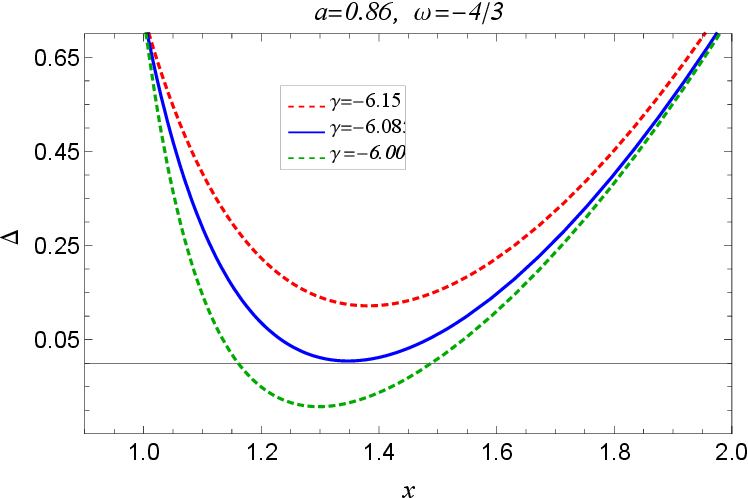}&
			\includegraphics[scale=0.6]{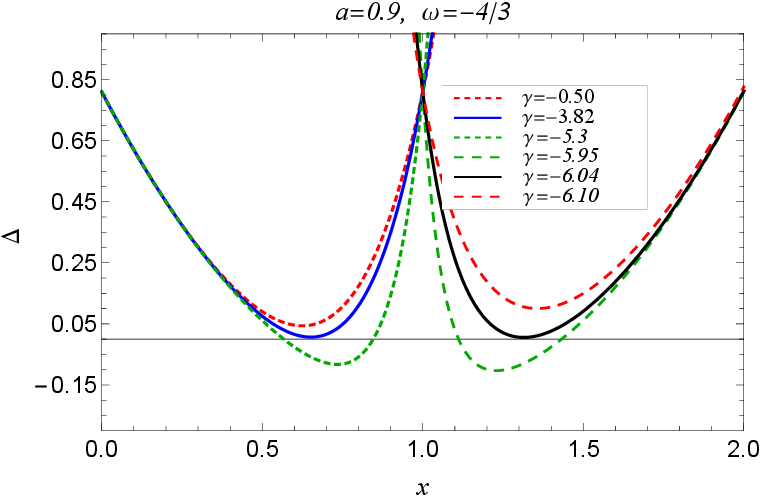}
		\end{tabular}
	\end{centering}
	\caption{The parameter space of $a$ and $\gamma$ at $\omega=-4/3$ (top right). The behaviour of horizons at $\omega=-4/3$ with varying black hole parameter $\gamma$ at (i) $a=0.60$ (top right) (ii) $a=0.86$ (bottom left) and (iii) $a=0.90$ (bottom right). The solid blue and black lines correspond to extreme values of parameters.}\label{plot3}		
\end{figure*} 

\begin{figure*}
	\begin{centering}
		\begin{tabular}{p{9cm} p{9cm}}
		    \includegraphics[scale=0.6]{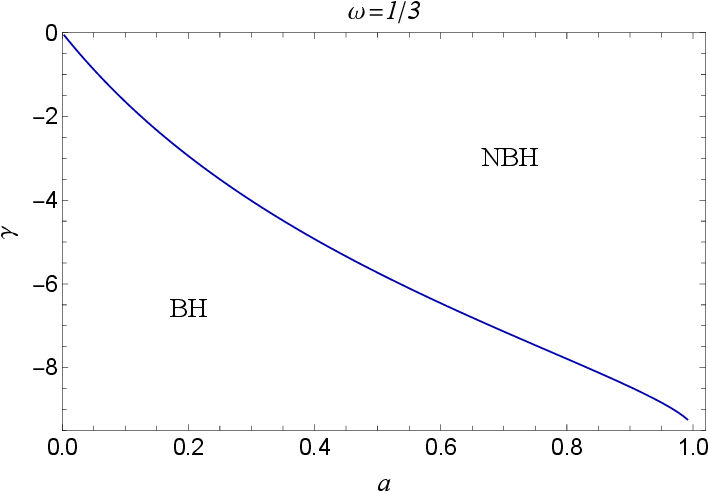}&
			\includegraphics[scale=0.6]{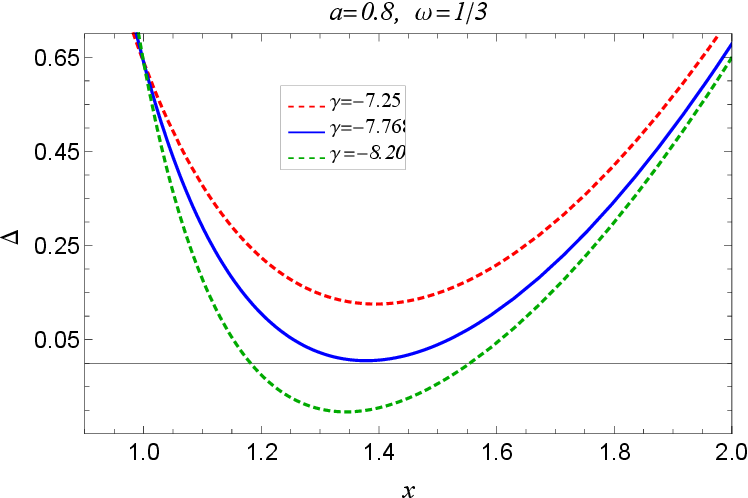}\\
		\end{tabular}
	\end{centering}
	\caption{The parameter space of $a$ and $\gamma$ at $\omega=1/3$ (top right). The behaviour of horizons at $\omega=-2/3$ and $a=0.80$ with varying black hole parameter $\gamma$. The solid blue line correspond to extreme value of parameter $\gamma$.}\label{plot4}		
\end{figure*} 
where $p_{t}(r)$, $\rho(r)$ and $p_{r}(r)$ are the tangential or transverse pressure, the energy density and the radial pressure of the fluid, respectively. The quantities $U_{\mu}$ and $N_{\mu}$ represent the four-velocity and radial unit vector, respectively,  
such that they obey the conditions $U_{\nu}U^{\nu}=1$, $N_{\nu}N^{\nu}=-1$ and $U_{\nu}N^{\nu}=0$. The matter Lagrangian density associated with the anisotropic fluid is given by $L_{m}=(-1/3)(p_r+2p_t)$ \cite{Deb:2018sgt}. Ghosh \cite{Ghosh:2015ovj} was the pioneer in obtaining the rotating counterpart of Kiselev black holes. Eq. (\ref{e4}) can be written as		
\begin{equation}\label{eqt16}
    \Theta_{\mu\nu}=-2T_{\mu\nu}-\frac{1}{3}(p_r+2p_t)g_{\mu\nu}.
\end{equation}
On using Eq's (\ref{e7}), (\ref{anytensor}) and (\ref{eqt16}), a  static spherical symmetric solution of Einsteins equations reads
\begin{eqnarray}\label{NR}
ds^2 &=& \left(1-\frac{2M}{r}+\frac{K}{r^{d}}\right)dt^2 
-\frac{dr^2}{\left(1-\frac{2M}{r}-\frac{K}{r^{d}}\right)}-r^2 d\Omega^2,
\end{eqnarray}
where
\begin{equation}
    d = \frac{8 (\gamma  \omega +\pi  (3 \omega +1))}{\gamma  (3-\omega )+8 \pi }.
\end{equation}
Thus we have a general form of exact spherically symmetric black holes in $f(R,T)$ gravity describing a Kiselev black hole, i.e., a black hole solution surrounded by the quintessence matter. The parameter $-1<\omega<-1/3$ for a de Sitter horizon causes acceleration, and $-1/3<\omega<0$ for an asymptotically flat solution. Here, $\omega$ is the parameter of the equation of state, $\gamma$ is the model-dependent parameter from the $f(R,T)$ gravity, $K$ and $M$ are integration constants. The Kiselev black hole reduces to Schwarzschild black hole for $K=0$, and to Reisner-Nordstrom black hole for $d=2$ and $K=Q^2$.

\section{Rotating Black hole}\label{sec3}
The Newman$-$Janis algorithm, initially designed within the framework of general relativity, has found widespread application in building rotating black hole solutions from their non-rotating counterparts \cite{Newman:1965tw}. Recently, this algorithm has also been utilized in modified gravity theories to generate rotating solutions based on non-rotating configurations \cite{Johannsen:2011dh,Jusufi:2019caq,Bambi:2013ufa,Ghosh:2014hea,Moffat:2014aja}. However, it is essential to note that when applying the Newman$-$Janis algorithm to arbitrary non-general relativity spherically symmetric solutions, specific issues may arise in the resulting axially-symmetric metric \cite{Hansen:2013owa}. In rotating black holes derived from modified gravity using the Newman$-$Janis algorithm, additional sources will probably arise alongside the original ones.  The metric for a rotating black hole in Einstein-Gauss-Bonnet gravity is also generated by utilizing the modified Newman$-$Janis algorithm, which incorporates Azreg-A\"\i{}nou's non-complexification procedure \cite{Azreg-Ainou:2014pra,Azreg-Ainou:2014aqa}. This procedure has successfully generated other rotating solutions with imperfect fluid content in Boyer$-$Lindquist coordinates, starting from spherically symmetric static solutions \cite{Kumar:2023jgh,Kumar:2022vfg,Brahma:2020eos,Islam:2022wck,Afrin:2022ztr,Kumar:2020hgm,Kumar:2020owy,KumarWalia:2022aop} and it can also generate  rotating Kislev black hole solutions \cite{Ghosh:2015ovj}.  The resulting  rotating black hole metric,  the counterpart of the spherically symmetric solution (\ref{NR}), governed by parameters $M$,  $a$, $\omega$, and $\gamma$  encompasses the Kerr solution \cite{Kerr:1963ud} and also generalizes the Kerr-Newman solution \cite{Newman:1965my}, which in Boyer-Lindquist coordinates reads 
\begin{align}
ds^2=&\left(\frac{\Delta - a^2 \sin^2 \theta}{\Sigma}\right) dt^2 - \frac{ \Sigma}{\Delta }  \, dr^2  \nonumber \\
& + 2 a \sin^2 \theta \left(1 - \frac{\Delta - a^2 \sin^2 \theta}{\Sigma} \right) dt \, d \phi
- \Sigma \, d \theta^2
\nonumber \\
&  -  \, \sin ^2 \theta  \left[ \Sigma + a^2 \sin^2 \theta \left(2 - \frac{\Delta - a^2 \sin^2\theta}{\Sigma}\right)   \right]    d \phi^2,\label{rotbhtr}
\end{align}
with 
\begin{eqnarray}
\Delta &=& r^2+a^2-2 r M(r),\quad \Sigma=r^2+a^2\cos^2\theta, \nonumber\\
M(r) &=& M-\frac{K}{2r^{d-1}},    
\end{eqnarray}
with $a$  being the  spin parameter. The metric Eq.~(\ref{rotbhtr}) reverts to Kerr black holes for the special case  $K\to 0$, to Kerr-Newman black holes when $K=Q^2$ and $d=2$ or when $\gamma$ and $\omega$ are related via
\begin{equation}
    \gamma = \frac{-4\pi(3\omega-1)}{5\omega-3}
\end{equation}
and to spherically symmetric black holes (\ref{NR}) when only $a=0$. For definiteness, we call the five-parameter metrics~(\ref{rotbhtr}) $-$ the $f(R,T)$ gravity-motivated rotating Kiselev black holes (FRKBH). To comprehensively analyse the FRKBH, we employ specific parameter values $w$ corresponding to black holes surrounded by various fields. These encompass dust ($w=0$), radiation ($w=1/3$), quintessence field ($w=-2/3$), and phantom field ($w=-4/3$) \cite{Kiselev:2002dx,Ghosh:2015ovj}. 
Interestingly, similar to the Kerr spacetime, the FRKBH spacetime metric (\ref{rotbhtr}) nevertheless maintains the time-translational and rotational invariance isometries, which, respectively, entail the existence of two Killing vector fields $\eta^{\mu}_{(t)}=\left(\frac{\partial}{\partial t}\right)^{\mu} $ and $\eta^{\mu}_{(\phi)}=\left(\frac{\partial}{\partial \phi}\right)^{\mu}$.

\begin{figure*}
	\begin{centering}
		\begin{tabular}{p{9cm} p{9cm}}
		    \includegraphics[scale=0.75]{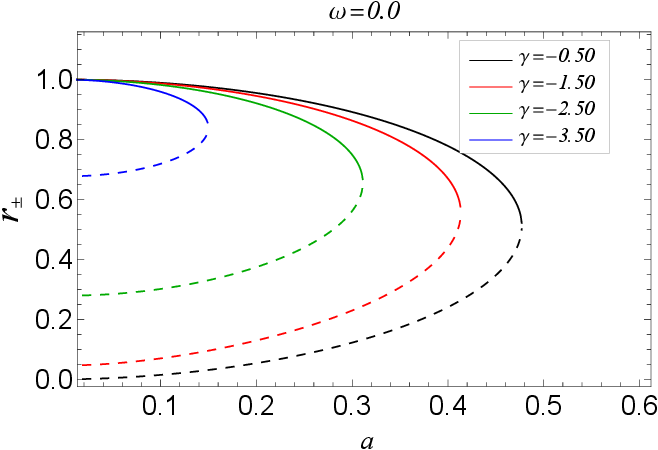}&
			\includegraphics[scale=0.75]{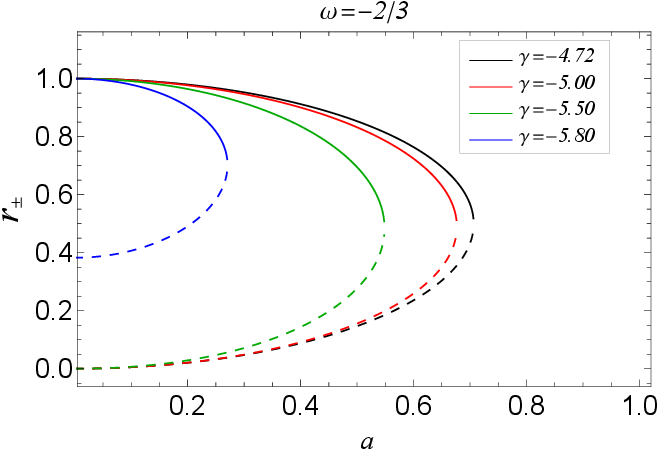}\\
            \includegraphics[scale=0.75]{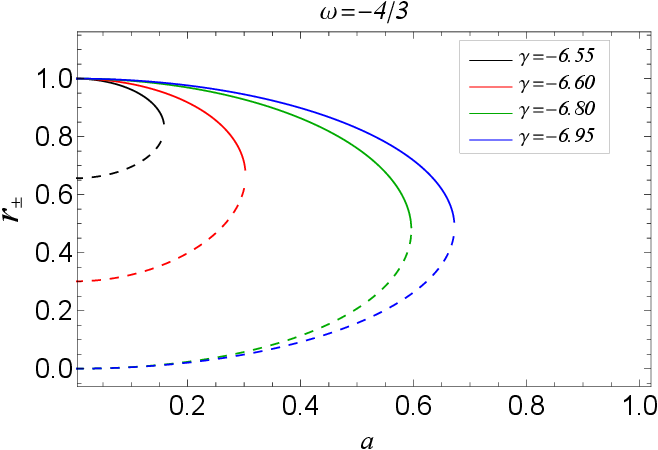}&
			\includegraphics[scale=0.75]{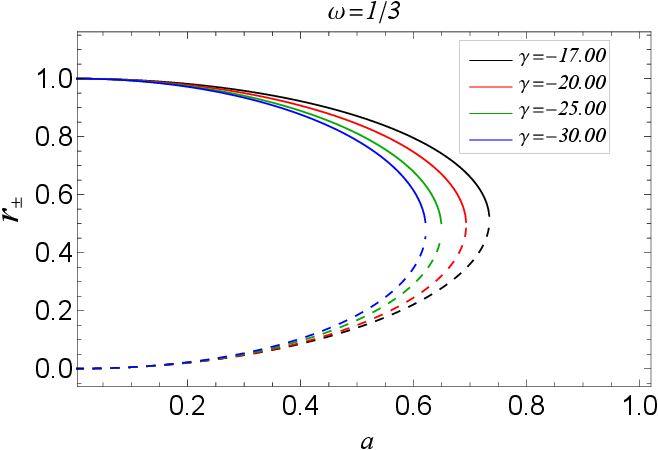}
		\end{tabular}
	\end{centering}
	\caption{Behaviour of event horizon (solid lines) and Cauchy horizon (dashed lines) vs spin parameter $a$ for different values of $\gamma$ and $\omega=0,~-2/3,~-4/3,~1/3$. }\label{plot5}		
\end{figure*} 

\begin{figure*}
	\begin{centering}
		\begin{tabular}{p{9cm} p{9cm}}
		    \includegraphics[scale=0.65]{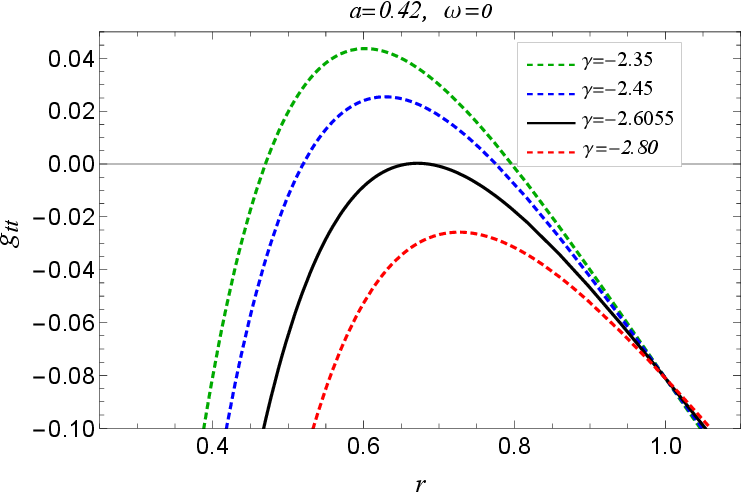}&
			\includegraphics[scale=0.65]{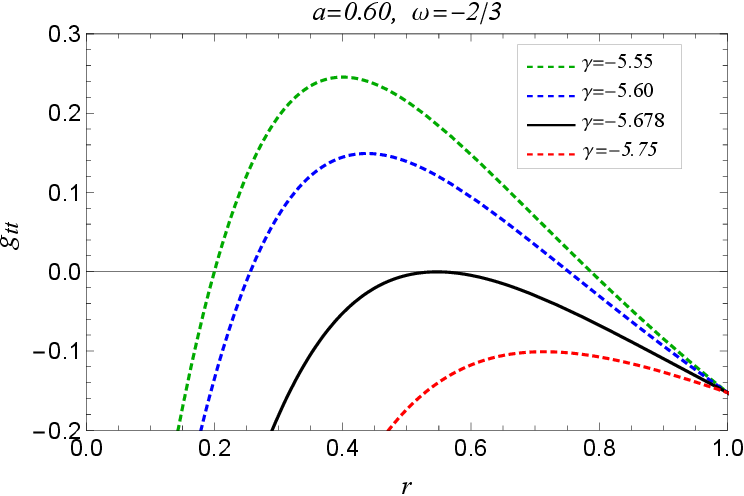}\\
            \includegraphics[scale=0.65]{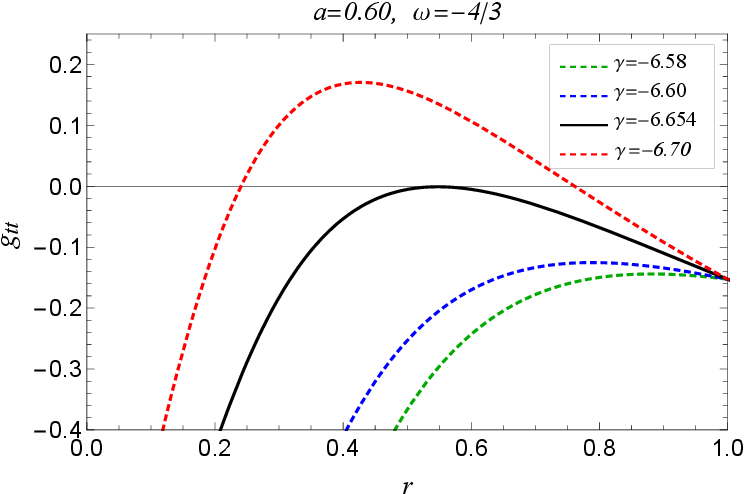}&
			\includegraphics[scale=0.65]{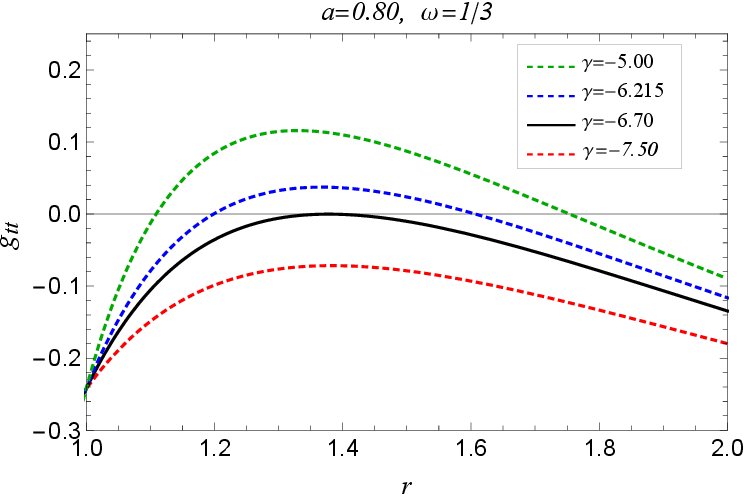}
		\end{tabular}
	\end{centering}
	\caption{The behavior of the SLS with varying spin $a$ and $\gamma$ for  $\omega=0,~-2/3,~-4/3,~1/3$. }\label{plot5b}		
\end{figure*} 
In order to further analyze the source associated with the metric (\ref{rotbhtr}), we use an orthonormal basis in which the energy momentum tensor is diagonal  \cite{Bambi:2013ufa,Neves:2014aba,Ghosh:2014pba}
\begin{equation}
	e^{(a)}_{\mu}=\left(\begin{array}{cccc}
		\sqrt{\mp(g_{tt}-\Omega g_{t\phi})}& 0 & 0 & 0\\
		0 & \sqrt{\pm g_{rr}} & 0 & 0\\
		0 & 0 & \sqrt{g_{\theta \theta}} & 0\\
		{g_{t\phi}}/{\sqrt{g_{\phi\phi}}} & 0 & 0 & \sqrt{g_{\phi\phi}}
	\end{array}\right),\label{Matrix}
\end{equation}
with $\Omega= g_{t\phi} /{g_{\phi\phi}}$. The components of the energy momentum tensor in the orthonormal frame read
\begin{equation}
	T^{(a)(b)} = e^{(a)}_{\mu} e^{(b)}_{\nu} G^{\mu \nu}. \nonumber
\end{equation}
Considering the line element (\ref{rotbhtr}), we can write the components of the respective energy momentum tensor as
\begin{eqnarray}
\rho &=& \frac{(d-1)K}{(a^2+r^2)^2 r^{d-2}} =-P_1, \nonumber \\
P_2 &=& \frac{(d-1)\Big[a^2(d+2)+dr^2\Big]}{2(a^2+r^2)^2 r^{d}} = P_3,
\end{eqnarray}
\subsection{Energy conditions and horizons }
To check the weak energy condition, we can choose an appropriate orthonormal basis \cite{Bambi:2013ufa,Neves:2014aba,Ghosh:2014pba} in which the energy momentum tensor reads
\begin{equation}
	T^{(a)(b)} = \mbox{diag}(\rho, P_1,P_2,P_3).
\end{equation}
The weak energy condition requires $\rho\geq0$ and $\rho+P_i\geq0$ ($i=1,\;2,\;3$) \cite{Zaslavskii:2010qz}. Therefore, $K\ge0$ and  $d\ge1$,  which are assumptions used throughout this work, are necessary for the weak energy condition to be satisfied.   Furthermore, the strong energy condition is given by 
\begin{equation}
\rho\geq0, \quad  \rho+P_i\geq0  (i=1,\;2,\;3), \quad \rho + P_{r}+2 P_{\theta} \geq 0,
\end{equation}
which are also fulfilled only when $K\ge0$ and  $d\ge1$.

The event horizon is  a stationary null surface that serves as the origin of null geodesic rays that are projected into the future but are never able to travel arbitrarily far from the black hole \cite{Hawking:1971vc,Poisson:2009pwt}, are defined by the surfaces $g^{\mu\nu}\partial_{\mu}r\partial_{\nu}r=g^{rr}=\Delta=0,$ and thus, the radii are zeros of
\begin{equation}
r^2+a^2-2Mr+\frac{K}{r^{d-2}}=0.\label{horizon}
\end{equation}
For the special case $d=2$, and $K=Q^2$, Eq.~(\ref{horizon}) reduces to
\begin{equation}
r^2+a^2-2Mr+Q^2=0,\label{horizonKN}
\end{equation}
where $K$ is identified as the charge $Q^2$, and solutions of the above equation give radii of horizons for the Kerr-Newman black hole given by
\begin{equation}
r_{\pm}=M\pm\sqrt{M^2-a^2-Q^2}.
\end{equation} 
\begin{table*}[htb!]
\resizebox{18cm}{!}{
 \begin{centering}	
\begin{tabular}{|C{2cm}|C{2.5cm}|C{2cm}|C{2cm}|C{2cm}|}
$\omega$ & $a$ & $\gamma$ & $r_{+}$& $r_{-}$  \\ 
\hline\hline
\multirow{10}{*}{$0$} & \multirow{4}{*}{$a=0.42$} 
& $-0.50$    & 0.7517 & 0.2778    \\  \cline{3-5}
& & $-1.44$  & 0.5587 & 0.5587  \\  \cline{3-5}
& & $-5.74$  & 1.3450 & 1.3450  \\   \cline{3-5}
& & $-6.00$  & 1.5704 & 1.1549  \\   \cline{2-5}
& \multirow{2}{*}{$a=0.60$}
& $-6.36$   & 1.3954 & 1.3954  \\    \cline{3-5}
& & $-6.60$ & 1.6014 & 1.2109  \\    \cline{2-5}
&\multirow{4}{*}{$a=0.90$}  
& $-7.59$  & 1.3105 & 1.3105  \\    \cline{3-5}
& & $-7.85$  & 1.4256 & 1.331  \\  \cline{3-5}
& & $-9.40$  & 0.8411 & 0.5699  \\  \cline{3-5}
& & $-10.40$  & 0.6551 & 0.6551  \\
\hline\hline

\multirow{10}{*}{$-2/3$} & \multirow{4}{*}{$a=0.60$} 
& $-5.00$    & 0.7293 & 0.2565    \\  \cline{3-5}
& & $-5.38$  & 0.4825 & 0.4825  \\  \cline{3-5}
& & $-6.30$  & 1.3972 & 1.3972  \\   \cline{3-5}
& & $-6.35$  & 1.6071 & 1.2095  \\   \cline{2-5}
& \multirow{2}{*}{$a=0.80$}
& $-6.47$   & 1.3793 & 1.3793  \\    \cline{3-5}
& & $-6.60$ & 1.5897 & 1.1287  \\    \cline{2-5}
&\multirow{4}{*}{$a=0.90$}  
& $-6.60$  & 1.3149 & 1.3149  \\    \cline{3-5}
& & $-6.70$  & 1.4387 & 1.1077  \\  \cline{3-5}
& & $-7.20$  & 0.8734 & 0.5647  \\  \cline{3-5}
& & $-7.98$  & 0.6580 & 0.6580  \\
\hline\hline

\multirow{10}{*}{$-4/3$} & \multirow{4}{*}{$a=0.60$} 
& $-6.24$    & 1.5896 & 1.223    \\  \cline{3-5}
& & $-6.27$  & 1.3978 & 1.3978  \\  \cline{3-5}
& & $-6.805$  & 0.4872 & 0.4872  \\   \cline{3-5}
& & $-6.90$  & 0.6860 & 0.2891  \\   \cline{2-5}
& \multirow{2}{*}{$a=0.80$}
& $-6.08$   & 1.3498 & 1.3498  \\    \cline{3-5}
& & $-6.00$ & 1.5018 & 1.1364  \\    \cline{2-5}
&\multirow{4}{*}{$a=0.90$}  
& $-3.82$  & 0.6501 & 0.6501  \\    \cline{3-5}
& & $-5.30$  & 0.8507 & 0.5718  \\  \cline{3-5}
& & $-5.95$  & 1.4466 & 1.1121  \\  \cline{3-5}
& & $-6.04$  & 1.3160 & 1.3160  \\
\hline\hline

\multirow{2}{*}{$1/3$} & \multirow{2}{*}{$a=0.80$} 
& $-7.76$  & 1.3902 & 1.3902    \\     \cline{3-5}
& & $-8.20$  & 1.5644 & 1.1801  \\
\hline
\end{tabular}
\end{centering}
}	
	\caption{ Table summarizing the values Event horizons ($r_{+}$), Cauchy horizons ($r_{-}$) and extremal horizon $T_{+}=r_{-}$ for various $\omega$ and $\gamma$.}\label{table1}
\end{table*} 
An analysis of Eq.~(\ref{horizon}) reveals that it has a maximum of two real positive roots, corresponding to the inner Cauchy horizon ($r_2$) and outer event horizon ($r_1$), such that $r_2\leq r_1$. Two distinct real positive roots of $\Delta=0$ infer the nonextremal black hole, while no black hole in the absence of real positive roots of Eq.~(\ref{horizon}), i.e., no horizon exists. There exists a particular value of the parameter $\gamma$, $\gamma=\gamma_e$, for which an extremal black hole occurs, such that Eq.~(\ref{horizon}) admits a double root; i.e., the two horizons coincide $r_2=r_1=r_e$. We have explicitly shown that, for fixed values of $a$, $K$ and $\omega$, $r_1$ decreases and $r_2$ increases with increasing $\gamma$ and eventually coincide for the extremal value of $\gamma$, i.e., $r_2=r_1=r_e$ for $\gamma=\gamma_e$.  Moreover, we infer that it is possible to find extremal values of parameters $a=a_e$ for fixed $\gamma$ and $\omega$, and $\omega=\omega_e$ for fixed $a$ and $\gamma$, for which the algebraic equation $\Delta=0$ has double roots (cf. Figures~\ref{plot1}-\ref{plot4}).
\paragraph{Case $\omega=0,\; -2/3,\; \text{and}  -4/3:$}
Intrestingly, when $a<a_1$ (e.g., $a_1\approx0.5$ for $\omega=0$) and $a>a_2$ (e.g., $a_2=0.809814$ for $\omega=0$), there exists two extreme values of $f(R,T)$ gravity parameter viz. $\gamma=\gamma_{E1},\gamma_{E2}$ (or $\gamma_{E3},\gamma_{E4}$) such that $\Delta=0$ admits roots corresponding to extremal black holes with degenerate horizons, whereas for  $\gamma_{E2}<\gamma<\gamma_{E1}$, one gets an NS. However, when $a_1<a<a_2$, there is only one extreme parameter value, i.e., $\gamma_{E}$, corresponds to an extremal black hole with degenerate horizons. In this case, again, $\gamma<\gamma_{E}$ corresponds to the black hole with two horizons and $\gamma>\gamma_{E}$ leads to naked singularities.  
\paragraph{Case $\omega=1/3:$} In this case, for a given $a$, a critical value of $\gamma=\gamma_{E}$ exists, such that the $\Delta=0$ has a double root corresponding to an extremal black hole with degenerate horizons. When   $\gamma<\gamma_{E}$, $\Delta=0$ has two simple roots and has no zero for $\gamma>\gamma_{E}$. These two cases correspond to a non-extremal black hole with two horizons and an NS (or no-horizon spacetime). 

Further, the static observers in the stationary spacetime follow the worldline of the timelike Killing vector $\eta^{\mu}_{(t)}$, such that their four-velocity is $u^{\mu}\propto \eta^{\mu}_{(t)}$ with the proper normalization factor. These observers exist as long as $\eta^{\mu}_{(t)}$ is timelike, such that $\eta^{\mu}_{(t)}\eta_{\mu(t)}=g_{tt}=0$ or
\begin{equation}
r^2+a^2\cos^2\theta- 2Mr+\frac{K}{r^{d-2}}=0,\label{gtteq}
\end{equation}
defines the boundary of the static limit surface (SLS), which, apart from black hole parameters, also depends on $\theta$ and coincides with the event horizon only at the poles. For the particular case $K=Q^2$ and $d=2$, Eq.~(\ref{gtteq}) corresponds to the Kerr-Newman black hole as
\begin{equation}
r^2+a^2\cos^2\theta- 2Mr+Q^2=0\,label{gtteqKN}
\end{equation}
and admits the solutions
$$r_{SLS}^{\pm}=M\pm\sqrt{M^2-a^2\cos^2\theta-Q^2},$$
which can be identified as the SLS radii for the Kerr-Newman black hole. Equation (\ref{gtteq}) is solved numerically, and the behaviour of SLS is shown in Fig.~\ref{plot5b}. It is clear from Fig.~\ref{plot5b} that the radii of the SLS decrease with increasing $K$ and $a$. The two SLS, corresponding to the real positive roots of Eq.~(\ref{gtteq}), coincide for suitably chosen parameters. However, these extremal values are different from those for the degenerate horizons. For fixed values of $M$ and $a$, the SLS radii for the FKRBHs are smaller than the Kerr black hole values. Likewise, the Kerr black hole, apart from $\Delta=0$, which is merely a coordinate singularity, rotating metric (\ref{rotbhtr}) is also singular at $\Sigma=0$, which is attributed to a ring-shaped physical singularity at the equatorial plane of the centre of the black hole with radius $a$.
\section{Komar Mass and Angular Momentum}\label{sec5}
Moreover, zero angular momentum observers, which  are stationary observers with zero angular momentum concerning spatial infinity, but due to frame dragging, have the position-dependent angular velocity $\omega$ given by
\begin{equation}
\omega=\frac{d\phi}{dt}=-\frac{g_{t\phi}}{g_{\phi\phi}}=
\frac{a \left(a^2-\Delta +r^2\right)}{(a^2+r^2)^2-a^2\Delta \sin^2{\theta}},
\end{equation}
which increases as the observer approaches the black hole and eventually takes the maximum value at the event horizon:
\begin{equation}
\Omega=\left.\omega\right|_{r=r_1}=\frac{a}{(r_1^2+a^2)}\label{angvelocity},
\end{equation}
such that observers are in a state of co-rotation with the black hole. Here, $\Omega$ is the black hole angular velocity, which has the same form as the Kerr black hole \cite{Poisson:2009pwt,Chandrasekhar:1992}.

The conserved quantities associated with the asymptotically timelike and spacelike Killing vector fields, respectively, $\eta^{\mu}_{(t)} $ and $\eta^{\mu}_{(\phi)}$, correspond to the mass and angular momentum assigned to the stationary, asymptotically flat black hole spacetime. A general argument for equality of the conserved Arnowitt-Deser-Misner mass \cite{Arnowitt:1962hi} and of the Komar mass \cite{Komar:1958wp} for stationary spacetimes having a timelike Killing vector is established in Refs.~\cite{Jaramillo:2010ay,Shibata:2004qz}. Following the Komar \cite{Komar:1958wp} definitions of conserved quantities, we consider a spacelike hypersurface $\Sigma_t$, extending from the event horizon to spatial infinity, which is a surface of constant $t$ with unit normal vector $n_{\mu}$  \cite{Chandrasekhar:1992}. The two-boundary $S_t$ of the hypersurface $\Sigma_t$ is a constant $t$ and constant $r$ surface with unit outward normal vector $\sigma_{\mu}$. The effective mass reads \cite{Komar:1958wp}
\begin{equation}
M_{\text{eff}}=-\frac{1}{8\pi}\int_{S_t}\nabla^{\mu}\eta^{\nu}_{(t)}dS_{\mu\nu},\label{mass}
\end{equation}
where $dS_{\mu\nu}=-2n_{[\mu}\sigma_{\nu]}\sqrt{h}d^2\theta$ is the surface element of $S_t$, $h$ is the determinant of the ($2\times 2$) metric on $S_t$, and 
\begin{equation}
n_{\mu}=-\frac{\delta^{t}_{\mu}}{|g^{tt}|^{1/2}},\qquad \sigma_{\mu}=\frac{\delta^{r}_{\mu}}{|g^{rr}|^{1/2}},
\end{equation}
are, respectively, timelike and spacelike unit outward normal vectors. Thus, mass integral Eq.~(\ref{mass}) turns into an integral over closed 2-surface at infinity
\begin{align}
M_{\text{eff}}=&\frac{1}{4\pi}\int_{0}^{2\phi}\int_{0}^{\phi}\frac{\sqrt{g_{\theta\theta}g_{\phi\phi}}}{|g^{tt}g^{rr}|^{1/2}}\nabla^{t}\eta^{r}_{(t)}d\theta d\phi\nonumber\\
=& \frac{1}{4\pi}\int_{0}^{2\phi}\int_{0}^{\phi}\frac{\sqrt{g_{\theta\theta}g_{\phi\phi}}}{|g^{tt}g^{rr}|^{1/2}}\left(g^{tt}\Gamma^{r}_{tt}+g^{t\phi}\Gamma^{r}_{t\phi} \right)d\theta d\phi.
\end{align}
Using the metric elements Eq.~(\ref{rotbhtr}), we obtain the effective mass of the rotating Kiselev black hole
\begin{equation}\label{mass1}
M_\text{eff} = M - \frac{K}{2r^{d-1}} \left[\left(\frac{r}{a}+\frac{a}{r}\right)(d-1)\tan^{-1}\frac{a}{r}+1\right],
\end{equation}
which is corrected due to the field and goes over to the Kerr black hole case that is $M_{\text{eff}}=M$, when $K=0$. For the special case $d=2$ and $K=Q^2$, Eq.~(\ref{mass1}) resembles the effective mass for the Kerr-Newman black hole with $K$ as the electric charge $Q^2$ and reads \cite{Modak:2010fn}
\begin{equation}
M_{\text{eff}}^{KN}=M-\frac{Q^2}{2r^2a}\left[ (r^2+a^2)\tan^{-1}\left(\frac{a}{r}\right)+a r\right].
\end{equation}
The effective mass for the spherically symmetric Kiselev black hole ($a=0$) is obtained from Eq.~(\ref{mass1}) and reads
\begin{equation}
M_{\text{eff}}^{NR} = M-\frac{d~K}{2 r^{d-1}},   
\end{equation}
which reduces to the Reissner$-$Nordstrom black hole for $K=Q^2$ and $d=2$:
$$M_{\text{eff}}^{RN}=M-\frac{Q^2}{r},$$
and to the Schwarzschild black hole $M_{\text{eff}}^{S}=M$, when $K=0$.

\begin{figure*}
	\begin{centering}
		\begin{tabular}{p{9cm} p{9cm}}
		    \includegraphics[scale=0.65]{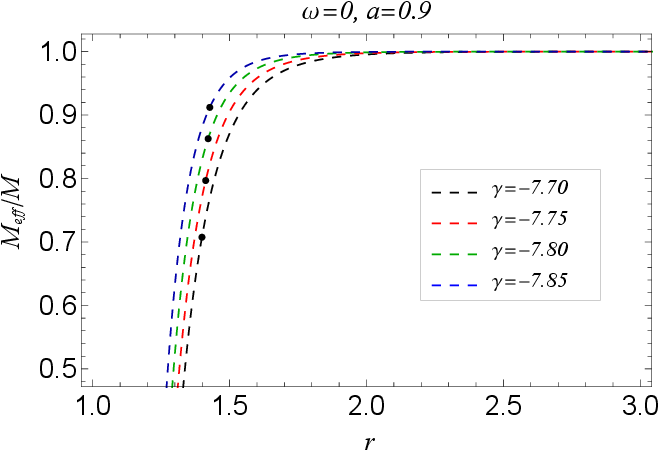}&
			\includegraphics[scale=0.65]{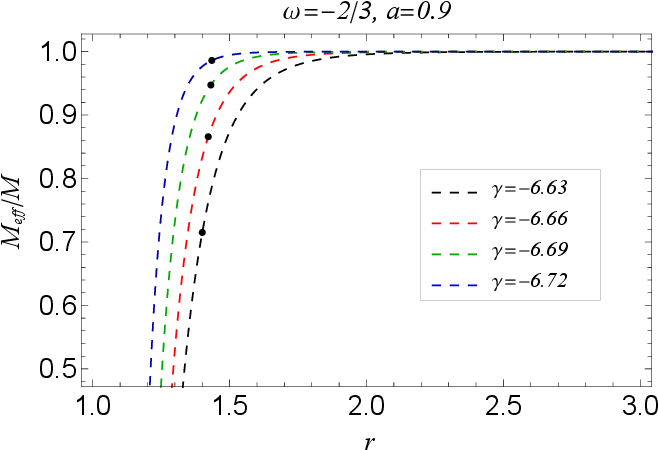}\\
			\includegraphics[scale=0.65]{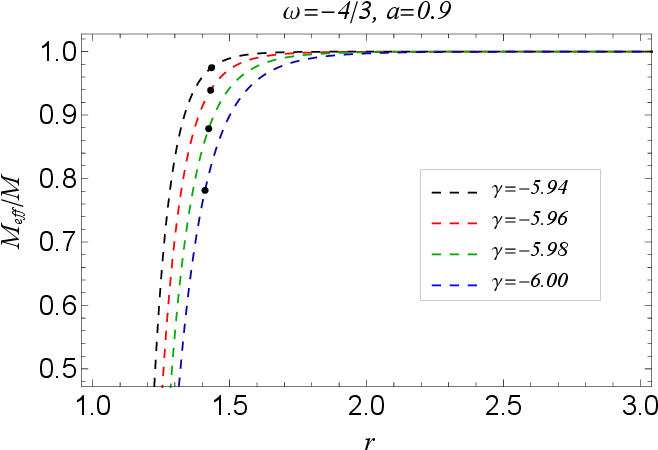}&
			\includegraphics[scale=0.65]{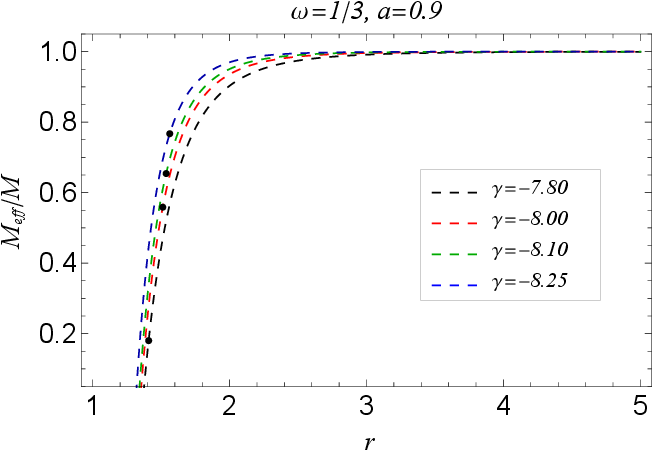}
		\end{tabular}
	\end{centering}
	\caption{The behaviour of effective mass with $r$ at $a=0.9$ and with varying $\omega$ and $\gamma$.}\label{plot6}		
\end{figure*}

\begin{figure*}
	\begin{centering}
		\begin{tabular}{p{9cm} p{9cm}}
		    \includegraphics[scale=0.65]{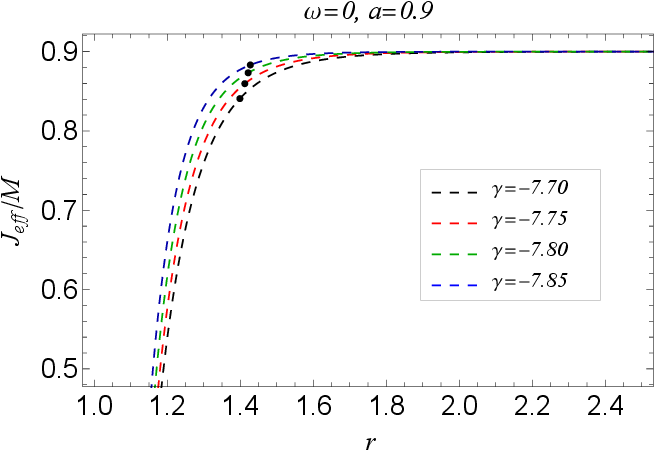}&
			\includegraphics[scale=0.65]{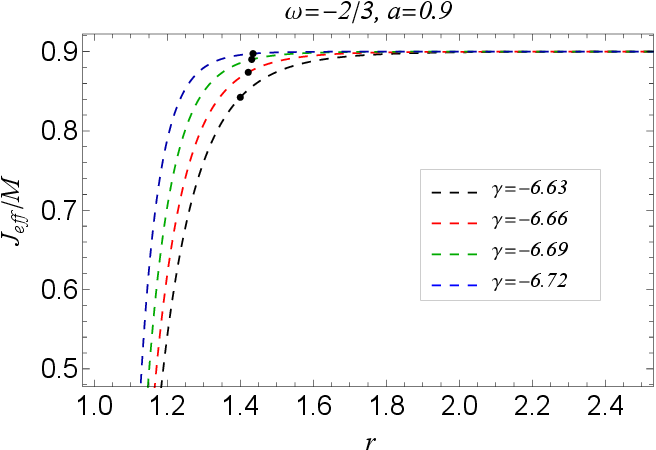}\\
			\includegraphics[scale=0.65]{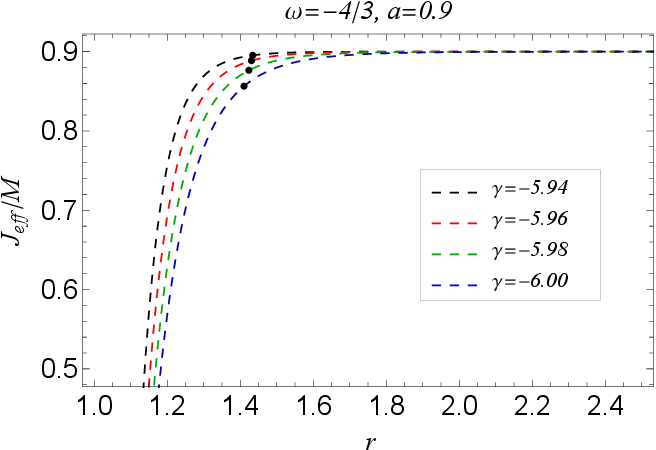}&
			\includegraphics[scale=0.65]{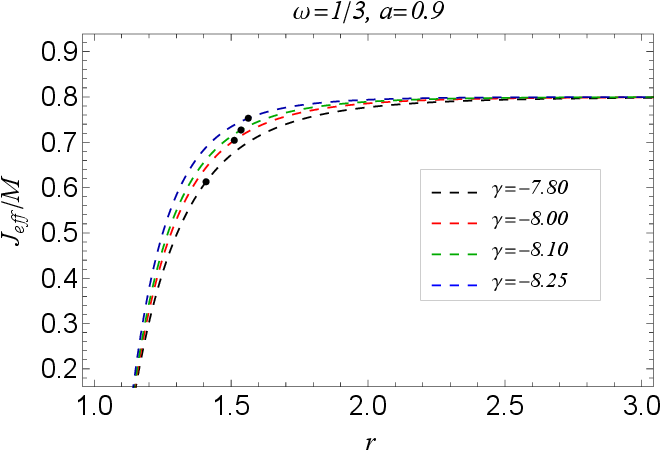}
		\end{tabular}
	\end{centering}
	\caption{The behaviour of effective angular momentum  with $r$ at $a=0.9$ and with varying $\omega$ and $\gamma$.}\label{plot7}		
\end{figure*}

Now, we use the spacelike Killing vector $\eta^{\mu}_{(\phi)}$ to calculate the effective angular momentum  \cite{Komar:1958wp}
\begin{equation}
J_{\text{eff}}=\frac{1}{16\pi}\int_{S_t}\nabla^{\mu}\eta^{\nu}_{(\phi)}dS_{\mu\nu},\label{ang}
\end{equation}
using the definitions of the surface element, Eq.~(\ref{ang}) recast as
\begin{align}
J_{\text{eff}}=&-\frac{1}{8\pi}\int_{0}^{2\phi}\int_{0}^{\phi}\nabla^{\mu}\eta^{\nu}_{(t)}n_{\mu}\sigma_{\nu}\sqrt{h}d\theta d\phi\nonumber\\
=& \frac{1}{8\pi}\int_{0}^{2\phi}\int_{0}^{\phi}\frac{\sqrt{g_{\theta\theta}g_{\phi\phi}}}{|g^{tt}g^{rr}|^{1/2}}\left(g^{tt}\Gamma^{r}_{t\phi}+g^{t\phi}\Gamma^{r}_{\phi\phi} \right)d\theta d\phi.
\end{align}
After performing the integration for the rotating Kiselev black hole Eq.~(\ref{rotbhtr}), this reads

\begin{eqnarray}\label{ang1}
J_{\text{eff}} &=& M a + \frac{K}{4r^{d}a^2} \Big[\left((d-3)a^2 + r^2 (d-1)\right)a r \nonumber \\ && - (d-1)(a^2+r^2)^2\tan^{-1}\frac{a}{r}    \Big] 
\end{eqnarray}
which  vanishes identically in the limiting case of $a=0$. For the particular case of  $d=2$ and $K=Q^2$ it reduces to
\begin{equation}
J_{\text{eff}}^{KN}=Ma+\frac{Q^2(r^2-a^2)}{4ar}-\frac{Q^2}{4a^2r^2}(r^2+a^2)^2\tan^{-1}\left(\frac{a}{r}\right),
\end{equation}
which can be identified as the Kerr-Newman black hole value \cite{Modak:2010fn}. In the asymptotic limits $r\to\infty$, the effective angular momentum Eq.~(\ref{ang1}) restores the value $J_{\text{eff}}^{K}=Ma$, which corresponds to the value for the Kerr black hole. Thus, the effects of the  field subside at a very large distance from the black hole. Equations (\ref{mass1}) and (\ref{ang1}) imply that at a finite radial distance the values of the effective mass and angular momentum get modified from their asymptotic values and depend on  of $\gamma$.  Figures \ref{plot6} and \ref{plot7}, graphically present the normalized values of the effective mass and angular momentum as functions of $r$ for various $\gamma$ values. It is evident from the figures that these effective quantities gradually decrease with decreasing $r$. Introducing parameters $\gamma$ and $\omega$ reduces the effective mass and angular momentum, specifically $M_{eff}/M\leq 1$ and $J_{eff}/Ma\leq 1$. Moreover, at a fixed radial coordinate, the normalized values of the effective quantities for regular black holes ($\gamma \neq 0$) are smaller than those for Kerr black holes ($K=0$) (See Ref. \cite{Kumar:2020hgm}). The impact of a non-zero $\gamma$ is significant only in the vicinity of the event horizon $r_+$. However, it diminishes at larger distances from $r_+$, resulting in $M_{eff}/M=1$ and $J_{eff}/Ma=1$ for large $r$ (see Fig's \ref{plot6} and \ref{plot7}). Consequently, $M_{eff}$ and $J_{eff}$ are consistently smaller than their asymptotic values.

It is well known that the Killing vectors $\eta^{\mu}_{(t)}$ or $\eta^{\mu}_{(\phi)}$ are not the  generators of the stationary black hole horizon; rather, it is their specific linear combination \cite{Chandrasekhar:1992}
\begin{equation}
\chi^{\mu}=\eta^{\mu}_{(t)}+\Omega \eta^{\mu}_{(\phi)},
\end{equation}
such that $\chi^{\mu}$ is globally timelike outside the event horizon, though it is a Killing vector only at the horizon \cite{Chandrasekhar:1992}. The Komar conserved quantity at the event horizon associated with $\chi^{\mu}$ reads as \cite{Komar:1958wp}
\begin{eqnarray}
J_{\chi}&=&-\frac{1}{8\pi}\int_{S_t}\nabla^{\mu}\chi^{\nu}dS_{\mu\nu}, \nonumber\\
&=&-\frac{1}{8\pi}\int_{S_t}\nabla^{\mu}\left( \eta^{\mu}_{(t)}+\Omega \eta^{\mu}_{(\phi)}\right)dS_{\mu\nu}.
\end{eqnarray}
Using Eqs.~(\ref{mass1}) and (\ref{ang1}), we obtain
\begin{eqnarray}
J_{\chi}&=&M_{\text{eff}}-2\Omega J_{\text{eff}},\nonumber\\
&=&\frac{M(r_1^2-a^2)}{(r_+^2+a^2)}-\frac{\left(r_+^2-(s-1)a^2\right)}{(r_+^2+a^2)s}\frac{K}{ r_+^{-(s-2)/s}}.\label{ST}
\end{eqnarray}
To understand the implication of the above conserved quantity, one must know the black hole horizon temperature \cite{Chandrasekhar:1992}
\begin{eqnarray}\label{temp}
T_+&=&\frac{\kappa}{2\pi}=\frac{\Delta'}{4\pi(r_+^2+a^2)},\nonumber\\
&=& \frac{r_+^2-a^2}{4\pi r_+(a^{2}+r_+^{2})} -\frac{K(d-1)}{4\pi r_+(a^{2}+r_+^{2})r_+^{d -2}}
\end{eqnarray}
whereas entropy is defined as follows
\begin{equation}
S_+=\frac{A}{4}=\pi(r_+^2+a^2).\label{entropy}
\end{equation}
Thus, we have derived the Komar mass and angular momentum for FKRBH spacetimes and calculated the other thermodynamical quantities. We can regain the exceptional cases of the Kerr and Kerr-Newman black holes in the proper limits. 
\begin{figure*}
	\begin{centering}
		\begin{tabular}{p{9.5cm} p{9.5cm}}
		    \includegraphics[scale=0.9]{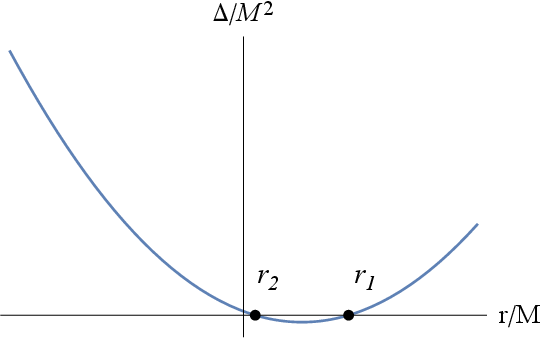}&
			\includegraphics[scale=0.5]{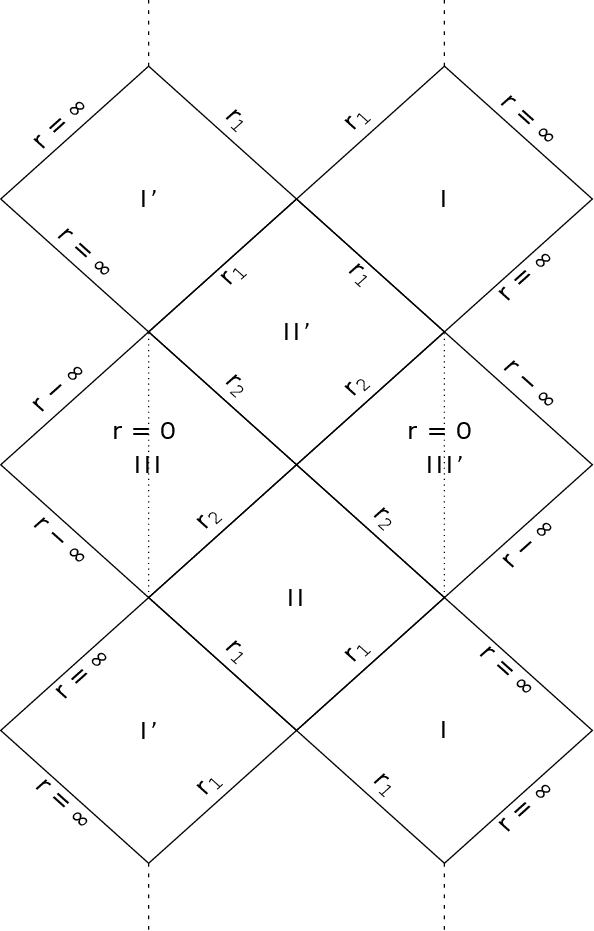}
		\end{tabular}
	\end{centering}
	\caption{\textit{Left:} Plot showing $\Delta(r)$ vs $r$  such that $\Delta(r)=0$ admits two positive root $r_2$ (event horizon) and $r_2$ (Cauchy horizon) at the values of $\omega$ and $\gamma$ for which $d=0,1,2$. \textit{Right:} Penrose diagrams of the corresponding spacetime  in the parameter space $(M,a,\omega,\gamma)$.  }\label{plot1a}		
\end{figure*} 

\begin{figure*}
	\begin{centering}
		\begin{tabular}{p{9.5cm} p{9.5cm}}
		    \includegraphics[scale=0.9]{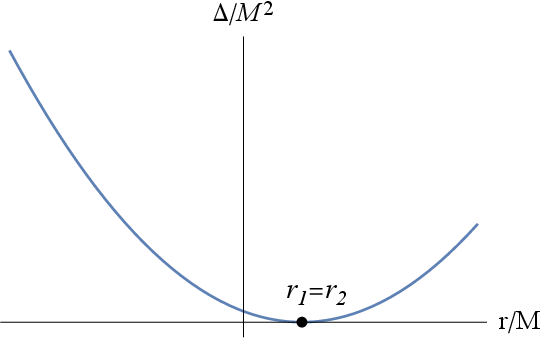}&
			\includegraphics[scale=0.5]{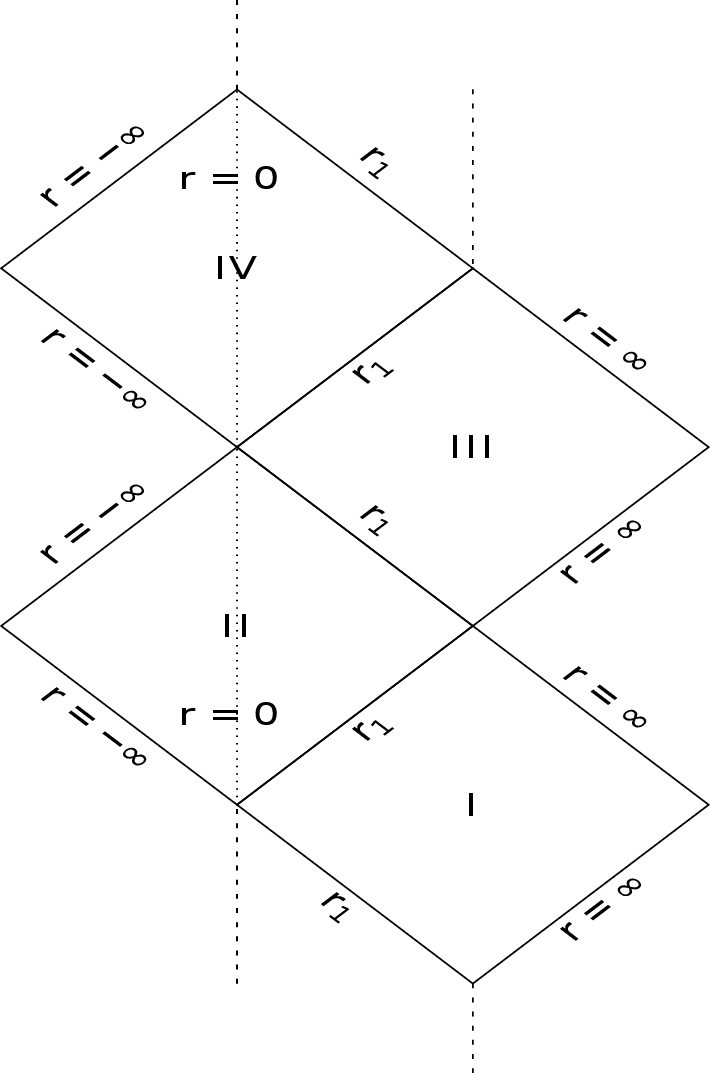}
		\end{tabular}
	\end{centering}
	\caption{\textit{Left:} Plot showing $\Delta(r)$ vs $r$  such that $\Delta(r)=0$ admits two positive equal roots $r_1=r_2$, corresponding to degenerate horizons at the values of $\omega$ and $\gamma$ for which $d=2$. \textit{Right:} Penrose diagrams of the corresponding spacetime  in the parameter space $(M,a,\omega,\gamma)$.}\label{plot1b}		
\end{figure*} 

\begin{figure*}
	\begin{centering}
		\begin{tabular}{p{9.5cm} p{9.5cm}}
		    \includegraphics[scale=0.9]{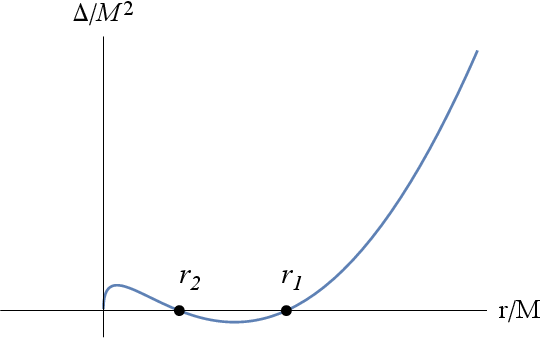}&
			\includegraphics[scale=0.5]{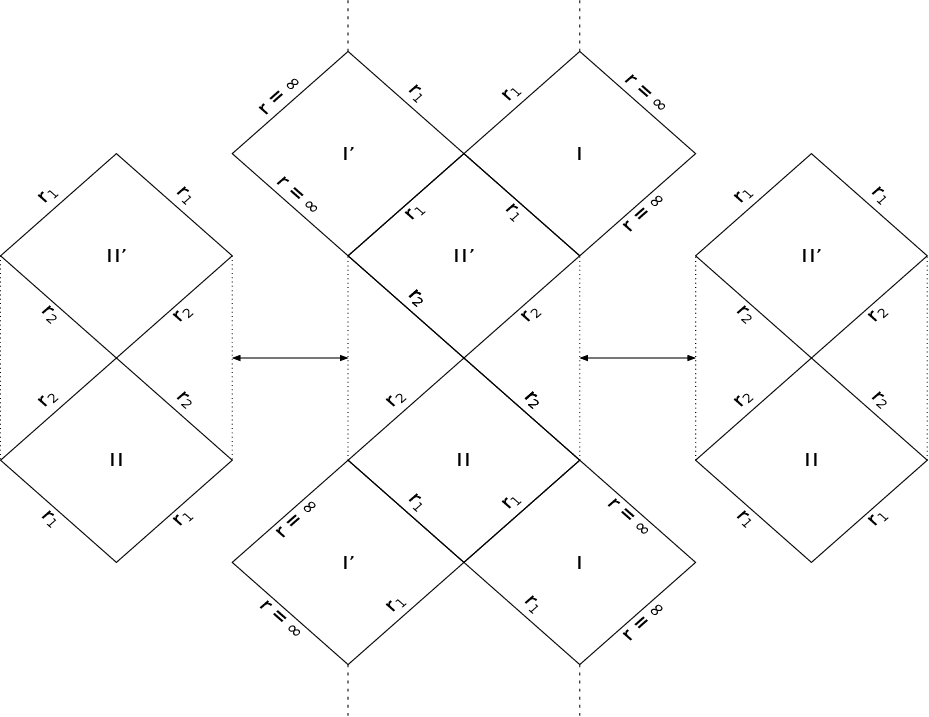}
		\end{tabular}
	\end{centering}
	\caption{\textit{Left:} Plot showing $\Delta(r)$ vs $r$  such that $\Delta(r)=0$ admits  two positive roots $r_1$ (event horizon) and $r_2$ (Cauchy horizon) at the values of $\omega$ and $\gamma$ for  for which $0<d<2$ and $d \neq 1$. \textit{Right:} Penrose diagrams of the corresponding spacetime  in the parameter space $(M,a,\omega,\gamma)$.}\label{plot1c}		
\end{figure*} 

\section{Structure of rotating spacetime: Penrose diagrams}\label{sec4}
The Penrose diagrams are designed to show the entire causal structure of any given geometry on a 2-D finite sheet. They provide a crucial road map for moving around a black hole.  Time is vertical, space is horizontal, and null rays are angled at 45$\degree$ to the axes, much like in the Minkowski diagram. A Penrose diagram, in contrast to a Minkowski diagram, has "points at infinity" added by compactification, which helps to visualize the intricate structure of infinity that results from the union of space and time. Along the symmetric $z$-axis ($\theta=0\degree$), we have generated diagrams for the FKRBH spacetime that correspond to each of the many points in the parameter spaces in Figs. \ref{plot1}-\ref{plot4}. 

For the case when the value of $\omega$ and $\gamma$ are such that $d=0,1$ or $2$, the Penrose diagrams, as well as the horizons, are precisely like that of Kerr black hole as given in Fig.~\ref{plot1a}. The surface $r=0$ is regular and time-like that exists well beyond inner horizon $r_1$. The Universe is represented by area I between $r=\infty$ and $r_1$, and the black hole is symbolized by region II between $r_1$ and $r_2$. Beyond the surface, $r=0$ to $r=-\infty$ lies the antiverse. The region I' is a mathematical extension of a second copy of the region I geometry glued along the anti-horizon in the opposite direction of time. In addition to a universe, antiverse, or black hole, the full analytic extension of the spacetime geometry also includes a parallel universe, a white hole, and a parallel antiverse as given by I', II', and part of region III' between $r=0$ and $r_{\infty}$, for the extremal black hole where the inner and outer horizon merge ($r_1=r_2$), the Penrose diagram is given in Fig.~\ref{plot1b} with $d=0,1$ or $2$. Its Penrose diagram illustrates that an extremal rotating Kiselev black hole has a different structure than a typical FKRBH. The spacetime, in this case, has only two regions, the region I between $r=\infty$ and $r_1$ is the  universe and the region II between $r_1$ and $r_{-\infty}$ which includes the timelike surface $r=0$. The region beyond $r=0$ is the antiverse. Finally, the Penrose diagram and the horizon structure for the case when $0<d<2$ but $d\neq1$ is shown in Fig.~\ref{plot1c}. In this case, the spacetime has a singularity at $r=0$ surface that lies beyond the inner horizon $r_1$.  

\section{Conclusion}\label{sec6}
Recently, Santos {\it et al.} \cite{Santos:2023fgd} made a significant discovery in $f(R,T)$ gravity. They investigated the parameter of the equations of state, denoted as $w$, and identified the first spherical black holes within this modified theory. This modification of the Kiselev black hole in the $f(R,T)$ gravity can reproduce well known solutions of the Einstein field equation as exceptional cases. They examined specific values of $w$ associated with black holes surrounded by various fields, such as dust ($w=0$), radiation ($w=1/3$), quintessence ($w=-2/3$), cosmological constant ($w=-1$), and phantom ($w=-4/3$).
In their study, the authors considered the model $f(T)=\varkappa T^{n}$ and explored the conditions necessary to generate Kiselev black holes in the context of $f(R,T)$ gravity. It was found that when $n$ takes on the accepted value of $1$, several particular values of the parameter $w$ in $f(R,T)$ gravity yield solutions that deviate from the Kiselev black hole in the general relativity.  However, it is challenging to test spherical black hole models through astrophysical observations alone, as black hole spin, precisely rotating black holes commonly observed in nature, play a crucial role in astrophysical processes. Testing modified gravity theories like $f(R,T)$ without incorporating rotating black hole models typically hinders observational verification.
The Kerr metric describes astrophysical black holes and remains the only stationary, vacuum axisymmetric metric that satisfies the Einstein field equations while avoiding pathologies beyond the event horizon. The groundbreaking observations made by the Event Horizon Telescope (EHT) provided images of supermassive black holes, namely Sgr A* and M87*. These observations, documented in studies such as \citep{EventHorizonTelescope:2019dse,EventHorizonTelescope:2019ggy,EventHorizonTelescope:2022xnr,EventHorizonTelescope:2022xqj}, revealed that the size of the black hole shadow agrees with predictions based on the Kerr metric, with an accuracy of around 10\%. This discovery offers an additional tool to investigate the nature of strong-field gravity and potentially constrain deviations from the Kerr metric, such as Kerr-like black holes that arise in theories like $f(R,T)$ gravity or other alternative gravity theories. Thus, the observations from the EHT can test fundamental theories of gravity, including $f(R,T)$ in the strong field regime, where a rotating black hole plays a vital role.
However, the scarcity of rotating black hole models in modified gravity significantly impedes progress in testing such theories through observational means. To address this limitation, we have tackled the problem by utilizing the revised Newman-Janis algorithm, a viable method for generating rotating solutions from a spherical Kiselev seed metric  (\ref{NR}).   The $f(R,T)$ gravity changes the structure of the Kerr black hole, resulting in the presence of an additional hair term in the metric (\ref{rotbhtr}). This metric describes FKRBH, which is asymptotically flat and encompasses various well-known black hole solutions, including Kerr ($K=0$), Kerr-Newman ($K=Q^2,\; d=2$), Reissner-Nordstrom ($K=Q^2,\; d=2,\; a=0$), and Schwarzschild ($K=0,\;a=0$) black holes.
Like the Kerr black hole, the FKRBH retains the Cauchy and event horizons and the stationary limit surface (SLS). However, the radii of these horizons and the SLS are affected by the parameter $\gamma$, resulting in their decrease. This has exciting implications for the ergosphere and can lead to significant consequences in the astrophysical Penrose process.
From the parameter space defined by $a$ and $\gamma$ for FKRBH, we can observe that when $a_1<a<a_2$, there is a critical value $\gamma=\gamma_E$ which corresponds to extreme value black holes characterized by degenerate horizons. When $a<a_1$ ($a>a_2$), we encounter two distinct critical values $\gamma=\gamma_{E1}, \; \gamma_{E2}$ with $\gamma_{E1}>\gamma_{E2}$ (or $\gamma=\gamma_{E3},\; \gamma_{E4}$ with $\gamma_{E3}>\gamma_{E4}$ (cf. Fig \ref{plot1}-\ref{plot4}). 

Despite the complexity of the FKRBH metric (\ref{rotbhtr}), we have analytically derived exact expressions for the conserved mass $M_{\text{eff}}$ and angular momentum $J_{\text{eff}}$ using the Komar prescription. These expressions hold at any radial distance. Notably, the $f(R,T)$ gravity introduces notable modifications to these conserved quantities compared to those of the Kerr black hole. However, in the limit $K=0$, the conserved mass $M_{\text{eff}}$ and angular momentum $J_{\text{eff}}$ converge to their values for the Kerr black hole. It is worth mentioning that the influence of the $f(R,T)$ gravity diminishes at large radial distances from the horizon, as $M_{\text{eff}}$ and $J_{\text{eff}}$ approach the values of the Kerr black hole in the asymptotic region ($r\to \infty$).

In conclusion, the resulting FKRBH spacetime (\ref{rotbhtr}) possesses mathematical properties reminiscent of the Kerr metric and several other intriguing features, demonstrating a rich spacetime structure.

Numerous avenues exist for future research, particularly in analyzing the accretion process onto black holes. Notably, the structure of Kerr black holes is significantly influenced by $f(R,T)$ gravity, which can have various astrophysical implications, such as the impact on wormholes, gravitational lensing properties, and black holes in AdS/CFT. Exploring the possibility of extending these findings to the other forms  of $f(R,T)$ gravity presents an exciting avenue for future investigations. In line with the no-hair conjectures, the role of FKRBH black hole deserves consideration and can potentially provide valuable insights. Moreover, studying the shadow of FKRBHs in light of results obtained from the observations made by the Event Horizon Telescope collaboration is essential.
\section{Acknowledgments} 
S.U.I and S.G.G. would like to  thank SERB-DST for project No. CRG/2021/005771. S.D.M acknowledges that this work is based upon research supported by the South African Research Chair Initiative of the Department of Science and Technology and the National Research Foundation.
\bibliography{KBH}
\bibliographystyle{apsrev4-1}
\end{document}